# Diboson production at LHC with warped extra dimensions

F. RICHARD[1]

Laboratoire de l'Accélérateur Linéaire,
IN2P2-CNRS et Université de Paris-Sud XI, Bât. 200, BP 34, 91898 Orsay Cedex France

**Abstract:** From the warped extra-dimensional model interpretation of the two forward-backward asymmetries observed on heavy quarks at LEP1, AFBb, and at Tevatron, AFBt, one predicts that LHC could observe, with the luminosity collected in 2011-2012, significant excesses in the diboson production for large invariant masses of the Z+W system, $m_{ZW}$, and, the W+W system, $m_{WW}$.

## 1. Introduction

Following our interpretations of the two anomalies observed on forward-backward asymmetry for b quarks AFBb at LEP1, ref1, and for top quarks AFBt at Tevatron, ref2, this note applies the Randal Sundrum (RS) phenomenology to the production of diboson pairs, ZW, WW and Zh, at LHC.
Recall that the excess on AFBt observed by both D0 and CDF, ref3, has triggered a large interest and received many different interpretations. While the production mode and the asymmetry are well defined at TeVatron, where 80% of the top production proceeds through valence quark anti-quark annihilation, life becomes more complicated at LHC where the main production mode is through gluon-gluon annihilation. Interpreting AFBt as due to a s-channel resonance with a mass between 1 and 2 TeV as in ref2, one may hope that LHC will observe an excess in that mass region where the production mechanism receives significant contributions from quark anti-quark annihilations. It is however fair to say that at high masses it is difficult to separate energetic top jets from ordinary jets, the so-called 'boosted top' problem, which presently limits the sensitivity of LHC. No doubt that in a near future this problem will be very much improved.
 In the meanwhile, one can try to use other channels to confirm the RS origin of the LEP1/Tevatron anomalies. In this interpretation there is a Kaluza-Klein (KK) excitation of the gluon with a mass around 1.5 TeV and large coupling to left handed light quarks. In these models one also expects KK excitations of the Z and the W bosons at about the same mass. These particles can be produced at LHC by quark anti-quark annihilation. They decay preferentially into Higgs and Higgs-like particles like h, $Z_L/W_L$ and would therefore decay into $W_LW_L$, $Z_LW_L$, $Z_Lh$ and $W_Lh$. There would be no anomaly on $W_TW_T$ and $Z_TW_T$ final states which dominate the total cross section at low masses.
 As described in section 3, experimental selections can enrich the dibosons samples in longitudinal components, increasing the sensitivity to these new particles. Large enhancements are predicted in the RS model, compared to the Standard Model (SM), for $W_LW_L$ and $Z_LW_L$ cross sections for masses above 500-1000 GeV and therefore accessible with present Tevatron data. LHC would give direct access to the KK resonances.
The SM production mechanism of W pairs shows strong cancellations between photon, Z boson and quark exchanges and, in adding a RS component, one creates an unbalance which is greatly amplified at high WW masses.

---

[1] richard@lal.in2p3.fr

## 2. Theoretical framework

In the RS model, see ref4, one can assume that ordinary bosons, photons and W/Z bosons have KK excitations which couple preferentially to top quarks, Higgs and longitudinal W/Z bosons. The mechanism behind this enhancement has a 'geographical' origin. Higgs bosons and their partners $Z_L$ and $W_L$, which originate from Higgs symmetry breaking, are located near the IR brane. KK Z/W partners are also located preferentially near this brane, hence a significant enhancement of their couplings to $Z_L$ and $W_L$.

In this simplest version, one cannot fulfill the various indirect constraints without assuming KK particles much heavier than the EW scale. In order to avoid this 'little hierarchy' problem, two approaches are proposed. In the first approach, as in ref5, the SM symmetry $SU(2)_L \otimes U(1)$ is extended to avoid the LEP/SLD/Tevatron precision measurement constraints. For what concerns the present search, extended symmetry groups, of the type $SU(2)_R \otimes SU(2)_L \otimes U(1)$, introduce new heavy bosons called Z' and W' with masses almost equal to the KK masses. These extra symmetries allow KK masses down to ~1.5 TeV depending on detailed assumptions.

Another approach, as in ref6, is to modify the warped geometry in the region of the IR brane. In this case one can reduce the masses down to about 1 TeV without assuming extra symmetries and therefore any extra Z'/W' boson. Recall however that to explain the AFBb anomaly, as discussed in ref1, it is necessary to assume Z-Z' mixing as in the first approach.

In both approaches ZkkWW type vertices can only contribute to WW production through mixing effects given that, due to the orthonormalisation conditions within RS, only vertices containing two KK particles differ from zero (see ref 8). One can, for instance, consider that W mixes with Wkk with a mixing angle $s_{0L}$ which is typically proportional to $(m_W/m_{KK})^2$. If W is longitudinal, this mixing is reinforced by a RS factor $\xi \sim 6$. As will be discussed in section 4, the Z'WW vertex can provide additional contribution.

Given that, within RS, it is usually assumed that light quarks are weakly coupled to KK bosons and uncoupled to Z' bosons, one usually ends up thinking that, with the present luminosity, there is very little hope to measure any effect at LHC. There could fortunately be two favorable circumstances, suggested by our interpretation of the AFBt anomaly, which could save the day. Firstly, the size of this asymmetry implies that KK particles could have a mass below 2 TeV. This low mass, as discussed in ref2, is still compatible with precision measurements. One also needs to assume that light quarks, more precisely left-handed light quarks, couple more intensely than expected, about 3 time more, to KK particles, also implying an important coupling to Z'/W' bosons. Also, according to our interpretation of AFBb, Z'/W' could have a larger coupling constant than ordinary bosons. It is therefore worthwhile to carefully work out these various components and decide whether LHC is able to test soon RS effects in the diboson production. This is the main purpose of this paper.

As a by-product of this discussion, one can note that the $ZkkZ_Lh$ coupling goes like $\xi$ times the SM coupling with no suppression from a mixing factor. As discussed in section 5, there could also be a large reinforcement of the coupling from the $Z'Z_Lh$ contribution (see ref1). In practice, as we shall see, the visibility of these effects turns out to be marginal at LHC.

As discussed in Appendix, anomalous ZWW couplings (see ref7) can occur due to mixing effects alone. Mixing not only occurs at the ZWW vertex but also at the qqZ/W vertex. The WW process is primarily sensitive to deviation of the κz parameter while the ZW process is sensitive to the g1z parameter (see definitions in ref7).

To complete this brief theoretical discussion, one should recall that there are additional RS terms due to heavy quarks which are not taken into account in this note. Without these terms, the RS contributions would violate unitarity at very high energy, meaning that one is dealing with an effective theory. These terms could, similarly to what happens in the SM, partially cancel the s-channel contributions from the KK bosons. A full treatment of these effects is ultimately needed but will be ignored in the present estimates. One can only speculate that such effects would, by analogy to the SM, decrease the impact of RS on diboson production.

## 3. Experimental issues

At LHC, for low masses, only leptonic modes provide a pure sample of WW and ZW candidates with, for WW, two leptons and two neutrinos and, for ZW, three leptons, two of them forming a Z mass, and a neutrino. With missing neutrinos one cannot reconstruct precisely the observables needed to compute the quantities required to isolate the longitudinal modes. To do this, one needs to reconstruct the mass of the diboson system and the transverse momentum of the W/Z which, for a given mass, allows estimating the center of mass production angle θ. This angle goes like $\sin^2\theta$ for $W_L W_L / Z_L W_L$, while the $W_T W_T / Z_T W_T$ distribution is forward peaked due to the t-channel quark exchange (see figure 8 in Appendix). Also the decay angle distribution of a $W_L/Z_L$ goes like $\sin^2\theta^*$ while for a $W_T/Z_T$ it goes like $1+\cos^2\theta^*$. These features have two important consequences:
- Leptons originating from $W_T W_T / Z_T W_T$ tend to be forward peaked (large rapidity) and can be efficiently removed with a well chosen transverse momentum cut
- On the contrary leptons from $W_L W_L / Z_L W_L$ are centrally produced which results in a good acceptance

In the following, the assumed efficiency takes into account the branching ratios into leptons, eν and μν for W, ee and μμ for Z, and, for $W_L W_L / Z_L W_L$ at high invariant masses, a reconstruction efficiency of 40% for WW and 25% for ZW.

How well can one estimate the masses of the KK bosons? For ZW production, with only one missing neutrino, one can use transverse momentum balance to achieve a reasonable mass resolution (see ref10).For WW production, with two missing neutrinos, one can still define a transverse mass but with much poorer resolution. One can also use the transverse momenta of the two leptons which are of order $M_{WW}/4$ for $W_L W_L$. The lepton transverse momentum distribution from the KK resonance is shown in figure 3 in the next section and has indeed a large fraction of events with values above $M_{KK}/4$.

Can one use semileptonic final states? In doing so one would improve by almost an order of magnitude the detection efficiency. For the WW final state one would improve the mass determination since, as for ZW leptonic final states, one can reconstruct the missing neutrino. The QCD background, from W/Z+jets is large, but, at these very high masses, this background becomes manageable. In the following one will only consider the purely leptonic final states but one should Bear in mind that the sensitivity could be increased by adding the semileptonic modes.

## 4. The W$_L$W$_L$ case

From ref7, one can show that, for large s/m²z, the SM amplitude is simply given by:

$$M^q_{WLWL} = \frac{e^2}{4c_W^2}\sin\theta \left[-Q_q + I_{3q}(1-cot_W^2)\right]$$

where Qq is the quark charge, I3q the weak SM isospin and cotw is the cotangent of the Weinberg angle. This expression shows that the t-channel quark exchange can be neglected for large masses. Note that the angular distribution of the WW goes like sinθ which helps to separate it from the transversally polarized states which are forward peaked. With anomalous couplings and RS contributions, one has:

$$\Delta M^{qL/R}_{WLWL} = \frac{e^2 s}{4m_W^2}\sin\vartheta\left(\Delta\kappa_z(-Q_q + \frac{I_3}{s_W^2}) + s\sum\frac{Q_{qi}Q_{KK}}{s-m_{KK}^2}\right)$$

where there is a sum Σ on the heavy RS KK resonances, Zkk, γkk and Z'. The constants Qqi and Qkk are given in the Appendix. The latter expression shows that only the BSM (Beyond SM) terms are boosted by s/m²w while, due to cancelations which occur between the photon, the Z boson and the quark exchanges, the SM amplitude remains constant with energy. This is a key reason to explain why, in spite of the high mass resonances, large deviations could be observable at LHC. Near the resonance mass region, these propagators should be replaced by Breit Wigner shapes, as explained in the Appendix. With our choice of parameters, the Z' resonance turns out to be very wide, mainly due to the Zh channel. With Γ$_{Z'}$/M$_{Z'}$~1 the role of this resonance is attenuated for the WW channel.

The figure below gives the ratio of cross sections RS/SM for what concerns the W$_L$W$_L$ final states. It clearly shows that, with our choice of parameters, there is a precocious enhancement due to anomalous couplings. Present limits given by LHC and Tevatron are not incompatible with these anomalous couplings.

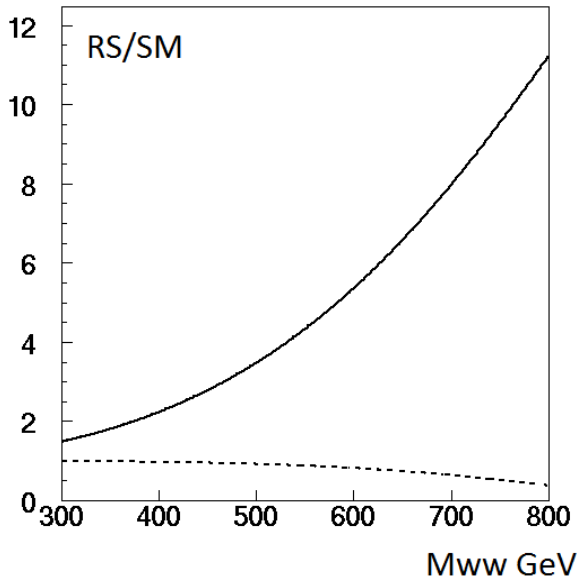

Figure 1: Ratio between the RS and the SM cross sections versus the WW invariant mass for the W$_L$W$_L$ process. The dotted curve is obtained by neglecting the anomalous couplings.

The dotted curve is obtained by neglecting the anomalous coupling which, as explained in the Appendix, is mainly due to mixing at the qqZ vertex. From this figure one also sees that an excess should become visible for masses above 500 GeV or, alternatively, for lepton transverse momenta above ~150 GeV (see the discussion in section 3). Note finally that this plot also applies to Tevatron and, from ref9, one can see that the CDF data already allow to explore lepton transverse momenta above 150 GeV.

The mass distributions shown in the two figures below correspond to the two RS scenarios described in section 2. An accumulated luminosity of 20fb-1 at 7 TeV and a reconstruction efficiency of 40% were assumed to draw these predictions.

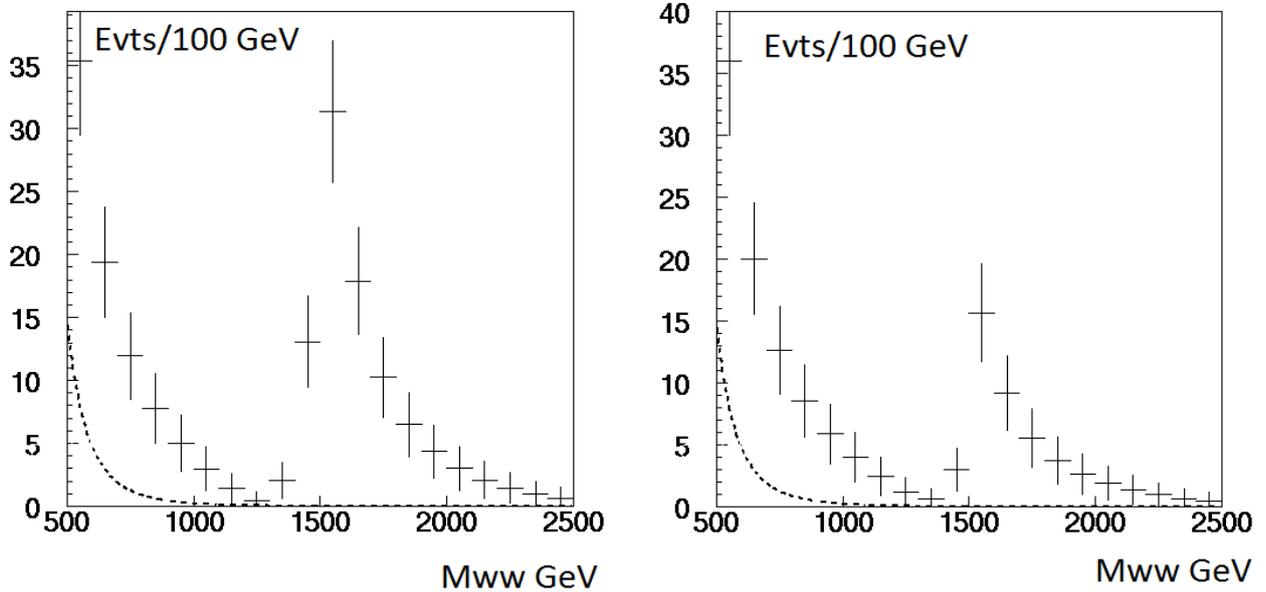

Figure 2: Number of $W_LW_L$ events per 100 GeV bin versus the WW invariant mass expected for 20fb-1 collected at LHC7. The left-hand side plot is with Z' while the right-hand is without this contribution. The dotted curves are the SM prediction for $W_LW_L$ events.

In practice, it will be difficult to reconstruct the WW mass with two missing neutrinos. As already commented, one benefits from the fact that the W are emitted at large pT and that the lepton tends to carry about half that pT. The figure above shows the pT distribution of leptons originating from the KK resonance which indeed peak at large values.

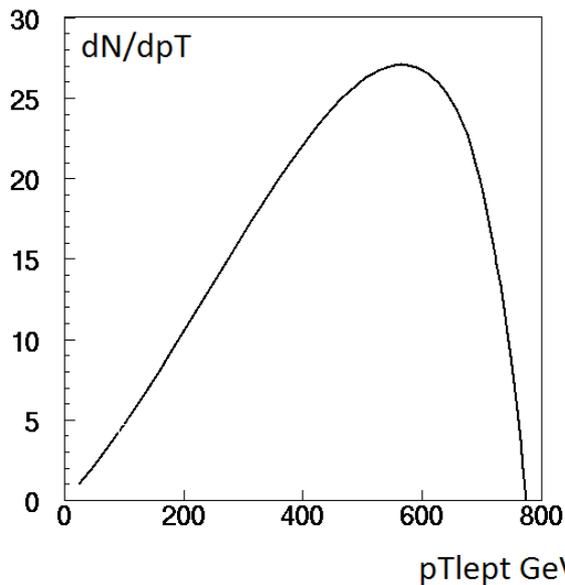

Figure 3: Lepton transverse momentum distribution originating from a 1.5 TeV KK resonance

## 5. The $Z_Lh$ case

For this channel one has direct $Z_{kk}Z_Lh$ coupling without mass mixing suppression and therefore there is, in principle, a large coupling.
At the **amplitude level**, one has:

$$\frac{RS}{SM} = 1 + \xi\left(s - m^2 z\right)\left[\frac{Q_{ZqL}\left(c_1 + \frac{g_R}{g_L} s_1 c' c_W\right)}{s - m_{KK}^2} + \frac{Q_{Z'qL}\left(s_1 - \frac{g_R}{g_L} c_1 c' c_W\right)}{s - m_{Z'}^2}\right]$$

The figure below displays the expected behavior showing, surprisingly, a weak enhancement.

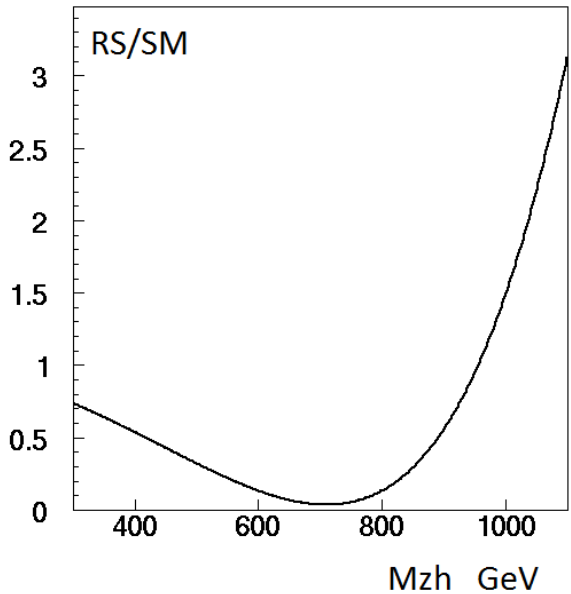

Figure 4: Ratio between the RS and the SM cross sections versus the Zh invariant mass for the $Z_Lh$ process.

Note that, with our choice of parameters, there is an accidental cancellation (recall that $s_1<0$) for the $Z_{KK}Z_Lh$ coupling which greatly reduces the cross section at resonance. This is not true for the Z'$Z_Lh$ coupling but recall that Z' is heavier and very wide and therefore has no great impact on the resonance.

What are the backgrounds to be considered?

1/ ZZ production gives similar final states unless one is able to resolve the Higgs and Z masses in b final states. One can however note that ZZ events are forward peaked due to quark exchange while h$Z_L$, for large h$Z_L$ masses, is distributed like sin²θ hence an obvious selection based on transverse momentum cut.

2/The process Z + jets has a huge cross section and requiring double b tagging may not suffice to reach an acceptable background level. Again mass reconstruction may not be precise enough to separate $Z_Lh$ from the continuum.

From above considerations, it follows that the RS effect could be very hard to observe in this channel.

## 6. The Z$_L$W$_L$ case

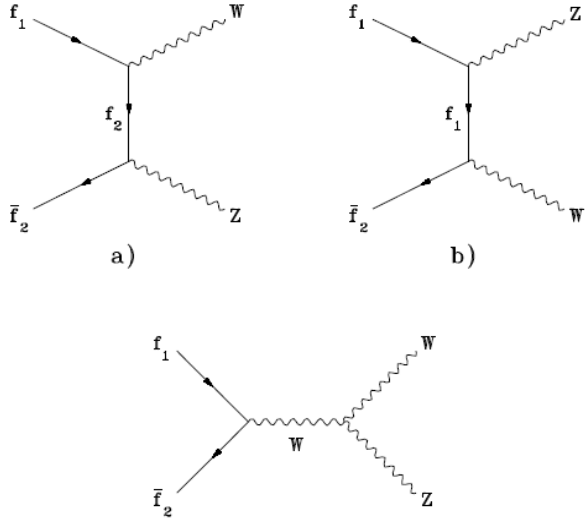

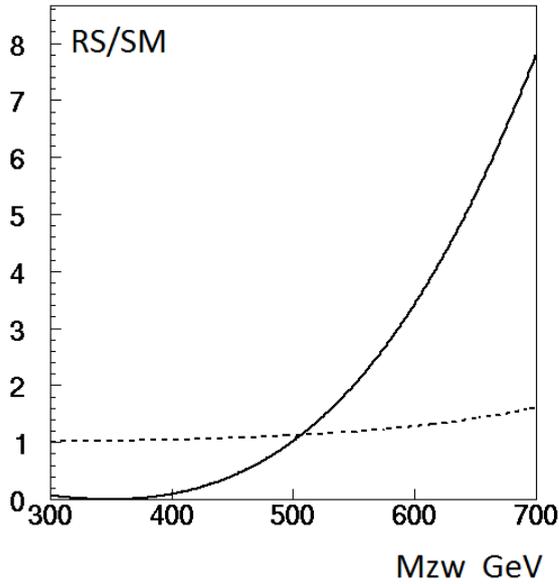

For this channel, the main process is q$_L$->Z$_L$W$_L$ with the contribution of the three diagrams shown above. Using formulae from ref7 and assuming that the ZW mass is large enough to have s/m²w>>1, the SM amplitude is given by:

$$M_{ZLWL} = -\frac{Q_W F}{2}\frac{m_Z}{m_W}\cot\theta_W \sin\theta$$

with F=Ce²/√2sinθ$_W$  C=δ$_{i1i2}$V$_{f1f2}$ where i1 (i2) is the color index of the incoming quark (anti-quark) and V$_{f1f2}$ is the quark mixing matrix element. Q$_W$ is the charge of the W.

Figure 5: Ratio between the RS and the SM cross sections versus the ZW invariant mass for the Z$_L$W$_L$ process (the dashed curve is without the anomalous coupling contribution).

For RS, for s/m²w>>1, one writes:

$$\Delta M_{ZLWL} \sim -\frac{s}{m_W m_Z}\left[\Delta g_Z^1 + s\Sigma\frac{Q^c_{qi} Q^c_{KK}}{s-m_{kk}^2}\right]\frac{F}{2}Q_W \cot\theta_W \sin\theta$$

This form clearly shows that for very large s, there is enhancement factor s/mzmw which greatly amplifies the effect of the KK resonances. The constants $Q^c_{qi}$ and $Q^c_{kk}$ are given in Appendix.

One can simply derive the $Z_L W_L$ cross section from the amplitudes above using the expression:

$$\sigma = \frac{1}{32\pi s}\int d\cos\theta |M_{LL}+\Delta M_{LL}|^2$$

Then one should convolute this cross section with the invariant luminosity u-dbar and d-ubar. One can neglect the contributions like u-sbar which are Cabbibo suppressed.

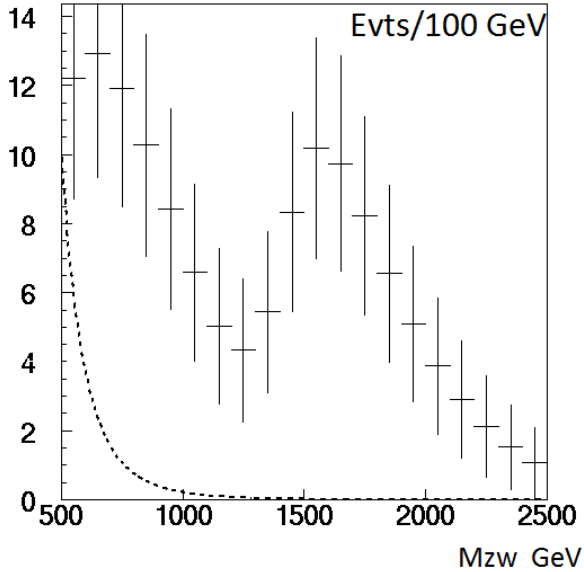

Figure 6: Number of $Z_L W_L$ events per 100 GeV bin versus the ZW invariant mass expected for 20fb-1 collected at LHC7. The dashed curve corresponds to the SM contribution for the longitudinal mode.

Above figure assumes 20fb-1 collected at LHC7. This prediction also assumes that WZ will be reconstructed through leptonic final states with a reconstruction efficiency of 25%. The dashed curve, which corresponds to the SM prediction for the longitudinal mode, is about half the total SM cross section in this high mass region. Ref11 indicates methods to enrich the resonance contribution allowing for a clear observation of an excess. Note also that smearing due to mass reconstruction has been ignored but one can expect a small effect given that the neutrino transverse momentum will be well reconstructed and given that the W is essentially transverse for this type of event.
The analysis presented in ref11, with 1fb-1, shows ZW candidates with masses up to 550 GeV with no excess with respect to the SM. So far this does not contradict significantly the RS prediction. With more luminosity one should observe candidates with masses above 1 TeV which would be unmistakable evidence for BSM physics.
The shape of the mass distribution given in figure 6 shows a resonance peaking around 1.5 TeV. As discussed in Appendix, the W' contribution is not observable since the W' is wide and has a 2 TeV mass.

## 7. Final comments

What could alter these very exciting predictions ? Changing moderately the KK parameters assumed in ref2 has large consequences. If one increases $M_{KK}$ from 1.5 to 2 TeV and assumes, at variance with ref1, that gz'~gz, the result shown in the figure below, with 50fb-1 collected at LHC7, indicates that no significant signal can be measured with 2012 data.

Finally, if such signals are observed at LHC, one may ask if there are alternative phenomenological interpretations aside from RS? An obvious candidate would be Technicolor (TC) scheme and its variants, with the presence of vector resonances (rho-type). Given that RS is claimed as dual to a composite model one may wonder if these two visions are really different. The only obvious feature which comes into mind is the prediction of light techni-pions which is clearly related to TC. If such particles are found by LHC, this would favor the TC interpretation. At Tevatron the evidence claimed by CDF, ref 12, was not confirmed by D0.

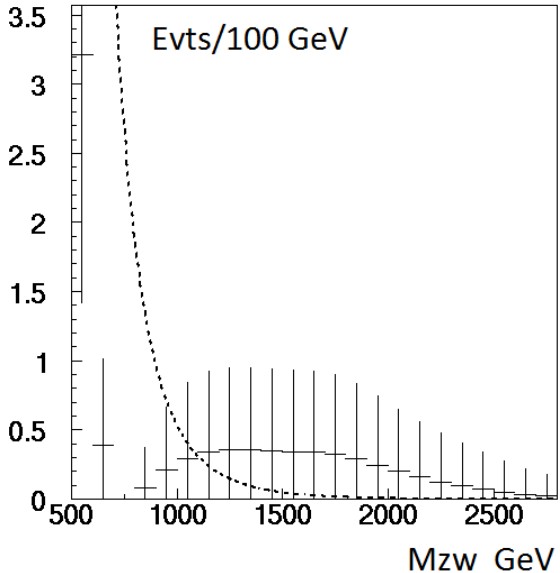

Figure 7: Number of $Z_LW_L$ events per 100 GeV bin versus the ZW invariant mass expected for 50fb-1 collected at LHC7 for mKK=2 TeV with gZ=gZ'. The dashed curve corresponds to the SM prediction for longitudinal modes.

## Conclusions

With the luminosity collected by LHC in 2011/2012, it becomes possible to achieve a meaningful test of the RS phenomenology in the diboson sector. If one assumes that the anomaly observed on AFBt at Tevatron is related to these models, one could not only observe an excess in of top anti-top around masses of 1.5 TeV but also measure a clear excess in ZW and WW events at large invariant masses. In a purely leptonic selection, which provides the best purity, ZW seems to offer the most promising prospect of observing a structure around 1.5 TeV.
For the WW channel, one would observe leptons with large transverse momenta clearly distinct from the SM background. Adding the semileptonic decays would greatly enhance the visibility of this channel and allow a better mass reconstruction of the WW resonance.
The RS model with custodial symmetry allows a variety of scenarios with different isospin assignments for the fermions and the eventual appearance of heavy quarks. This flexibility of the model prevents definite predictions signals at LHC. Even the favorable predictions provided by our scenario could be seriously degraded by assuming, for instance, that there is an additional RS heavy quark which would increase the total width of the resonance and/or by assuming moderate changes of the parameters deduced from LEP1/Tevatron anomalies.

# References

1. A. Djouadi et al. , Published in Nucl.Phys.B773:43-64,2007.
   e-Print: hep-ph/0610173
2. A. Djouadi et al. Published in Phys.Lett.B701:458-464,2011.
   e-Print: arXiv:1105.3158
   A. Djouadi et al., Published in Phys.Rev.D82:071702,2010.
   e-Print: arXiv:0906.0604
3. CDF collaboration (T. Aaltonen *et al.*), Published in Phys.Rev.D83:112003,2011.
   e-Print: arXiv:1101.0034
   D0 Collaboration (V. M. Abazov *et al.*), Published in *Phys. Rev. D 84, 112005 (2011)*.
   e-Print: arXiv:1107.4995
4. L. Randall and R. Sundrum, Phys. Rev. Lett. 83 (1999) 3370.
5. K. Agashe et al., Published in JHEP 0308:050,2003.
   e-Print: hep-ph/0308036
6. J. A. Cabrer et al., Published in JHEP 1105:083,2011.
   e-Print: arXiv:1103.1388
7. G. Gounaris *et al.* , e-Print: hep-ph/9601233
8. K. Agashe et al., Published in Phys.Rev.D80:075007,2009.
   e-Print: arXiv:0810.1497
   K. Agashe et al., Published in Phys.Rev.D80:075007,2009.
   e-Print: arXiv:0810.1497
9. CDF Collaboration Published in Phys.Rev.Lett.104:201801,2010.
   e-Print: arXiv:0912.4500
10. U. Baur et al., Published in Phys.Rev.Lett.72:3941-3944,1994.
    e-Print: hep-ph/9403248
11. CMS Collaboration CMS-PAS-EXO-11-041
    http://cdsweb.cern.ch/record/1377329
    ATLAS Collaboration (Georges Aad *et al.*).
    e-Print: arXiv:1111.5570
12. CDF Collaboration Published in Phys.Rev.Lett.106:171801,2011.
    e-Print: arXiv:1104.0699

**Acknowledgement:** Useful discussions with G. Hamel de Monchenault are gratefully acknowledged. Gregory Moreau, without whom this preliminary work would not have been possible, has very generously and patiently provided his precious expertise for the theoretical part.

# APPENDIX

In this Appendix various details are given allowing to understand which assumptions have been used to achieve the results given in the main text.

## $W_L W_L$

For large $s/m^2_z$, one has ($Q_q$ quark charge $I_3$ weak SM isospin).

$$M^{qL/R}_{WLWL} = \frac{e^2}{4c_W^2} \sin\theta \left[ -Q_q + I^q_{3L}(1 - cot^2_W) \right]$$

In the SM Qu=2/3, Qd=-1/3, I3L=0 for right handed quarks, I3L=1/2 for u quarks and I3L=-1/2 for d quarks. In RS one has an extra contribution given by:

$$\Delta M^{qL/R}_{WLWL} = \frac{e^2 s}{4m_W^2} \sin\vartheta \left( \Delta\kappa_z(-Q_q + \frac{I^q_{3L}}{s_W^2}) + s\sum \frac{Q_{qi}Q_{KK}}{s - m_{KK}^2} \right)$$

where:

$$\sum \frac{Q_{qi}Q_{KK}}{s - m_{KK}^2} = \frac{Q_{qL/RZ}Q_{ZKK}}{s - m_{ZKK}^2} - \frac{Q_{qA}Q_{AKK}}{s - m_{AKK}^2} + \frac{Q_{qL/RZ'}Q_{Z'}}{s - m_{Z'}^2}$$

Here one derives the couplings of Akk, Zkk and Z' to WW and quarks through mixing effects as in ref8. Various mixing effects are involved:
W-Wkk $s_{0L}$ ; W-W' $s_{0R}$ ; Z-Zkk $s_{01}$ ; Z-Z' $s_{01X}$ ; Zkk-Z' $s_1$ ; Wkk-W' $s^c_1$ ; W3-X $s'$ .
With our choice of parameters one has $s^c_1 \sim s_1$ as will be assumed in the following.
The table below shows 2 examples of solutions. The first solution, used in this paper, is motivated by our interpretation of the two AFB anomalies. From AFBt one takes $M_{kk}$ =1.5 TeV. From AFBb, one assumes a large gz' coupling. However, as compared to ref1, one reduces this coupling by a factor 2 given that Mkk is also reduced by the same factor.
The second solution, shown for comparison, comes from ref9.
From the expression of $s_{01X}$:

$$s_{01X} \sim -\xi \left(\frac{M_Z}{M_{Z'}}\right)^2 \left(\frac{g_{Z'}}{g_Z}\right)$$

one could infer that a large Z' coupling gives a large value for $Q_{Z'}$. However a large gz' also increases the Z' mass with respect to Mkk reducing the influence of the Z' propagator. This influence is further reduced by the larger width of the Z' resonance.

| $g_Z$ | $g_{Z'}$ | $g_L$ | $g_R$ | s' | $s_1$ | $s_{01}$ | $s_{01X}$ | $s_{0L}$ | $s_{0R}$ | $M_{Zkk}$ TeV | $M_{Z'}$ TeV | $Q_{zkk}$ | $Q_{z'}$ |
|---|---|---|---|---|---|---|---|---|---|---|---|---|---|
| 0.74 | 1.80 | 0.65 | 1.78 | 0.2 | -0.37 | 0.021 | -0.056 | 0.016 | -.025 | 1.5 | 2 | -.01 | -.05 |
| 0.74 | 0.74 | 0.65 | 0.65 | 0.55 | 0.48 | 0.013 | -0.01 | 0.01 | -0.01 | 2 | 2 | -.0013 | -.012 |

Note that for the first solution, used as a reference in this paper, all mixing angles are small, with the exception of s1, and one can therefore simplify the various expressions by retaining only the leading order terms. For instance in the case one has a combination of the type cisj, one can drop ci since $c_i \sim 1 - s_i^2/2$, which induces a 3d order term.
The coefficients $Q_{KK}$ used in the expression of $\Delta M_{WLWL}$ are given by:

$$Q_{Z'} = s_1 s_{01} + c_1 s_{01X} - 2s_1 s_{0L} \quad Q_{AKK} = -2s_{0L} \quad Q_{ZKK} = c_1 s_{01} - s_1 s_{01X} - 2c_1 s_{0L}$$

The coefficients $Q_{qi}$ used in the expression of $\Delta M_{WLWL}$ are given by:

$$Q_{AKKqL} = Q_q Q(c_{qL}) \qquad Q_{ZKKqL} = \left[c_1 s_{01} - s_{01X} s_1 + Q(c_{qL})c_1\right]\left(-Q_q + \frac{I_{3L}^q}{s_W^2}\right) - Q'(c_{qL}) s_1 \frac{g_{Z'}}{g_Z s_W^2}\left(I_{3R}^q - Y_q s'^2\right)$$

$$Q_{Z'qL} = \left[c_1 s_{01X} + s_{01} s_1 + Q(c_{qL})s_1\right]\left(-Q_q + \frac{I_{3L}^q}{s_W^2}\right) + Q'(c_{qL}) c_1 \frac{g_{Z'}}{g_Z s_W^2}\left(I_{3R}^q - Y_q s'^2\right)$$

where $Y_q = Q_q - I_{3L}^q$ and Q(cqL)=0.6 and Q'(cqL)=0.8.

In ref8 one assumes that Yq=Qq which is only legitimate when I3L=0.

For the left handed quarks one assume that:

$$I_{3R}^{uL} = I_{3R}^{dL} = 0$$

This follows from a detailed discussion given in ref1 and implies small couplings of ordinary quarks to Z' given the value chosen for s'. In contrast ref8 assumes that:

$$I_{3R}^{uL} = I_{3R}^{dL} = -\frac{1}{2}$$

This choice would imply much larger couplings to ordinary quarks which, in ref8, has no consequence since it assumes that Q'(cqL)=0 for ordinary quarks. In our case where Q'(cqL)=0.8 this choice for I3R would imply large effects. This remark shows that there could be important deviations from our predictions given the freedom of such models.

For right handed quarks one simply has:

$$Q_{ZKKqR} = -Q_q\left[c_1 s_{01} - s_{01X} s_1 + Q(c_{qR})c_1\right] \qquad Q_{Z'qR} = -Q_q\left[c_1 s_{01X} + s_{01} s_1 + Q'(c_{qR})s_1\right]$$

where Q(cqR)=-0.2 Q'(cqR)=0.

Again different isospin assignments are taken in ref1 and ref8 but this has no practical consequences since, in both models Q'(cqR)=0 and right handed light quarks do not couple to Z'.

## $Z_L W_L$

For RS one writes (assuming s/m²w>>1):

$$\Delta M_{ZLWL} \sim -\frac{s}{m_W m_Z}\left[\Delta g_Z^1 + s\Sigma \frac{Q_{qi}^c Q_{KK}^c}{s - m_{kk}^2}\right]\frac{F}{2} Q_W \cot\theta_W \sin\theta$$

with F=Ce²/√2sinθw  C=δ$_{i1i2}$V$_{f1f2}$ where i1 (i2) is the color index of the incoming quark (anti-quark) and V$_{f1f2}$ is the quark mixing matrix element. Q$_W$ is the charge of the W.

The coefficient Q$^c$$_{KK}$ and Q$^c$$_{qi}$ are given by:

$$Q_{WKK}^c = -s_{0R} s_1 - c_1 s_{01} \qquad Q_{WKKqL}^c = \left[-c_1 s_{0L} - s_1 s_{0R} + Q(c_{qL})c_1\right]$$

$$Q_{W'}^c = c_1 s_{0R} - s_{01} s_1 \qquad Q_{WqL}^c = \left[c_1 s_{0R} - s_1 s_{0L} + Q(c_{qL})s_1\right]$$

The amplitude for transverse modes is given by:

$$M_{ZTWT} = \frac{F}{\sin\theta}(\lambda_W - \cos\theta)\left[Q_W \cot\theta_W \cos\theta + \frac{\tan\theta_W}{3}\right]$$

where λ$_W$ =±1. For large transverse momentum there is no singularity at sinθ=0 such that Z$_T$W$_T$ and Z$_L$W$_L$ are of similar sizes which allows an easy observation of an excess due to the RS contributions.

The following plot, from ref7, illustrates the distinct angular distributions between the longitudinal mode (dotted curve) and the transverse mode (dashed curves). A cut on the transverse momentum of Z allows to get rid of the forward/backward peaks leading to a good purity signal. This plot also shows how an anomalous coupling modifies the angular distribution (full curve).

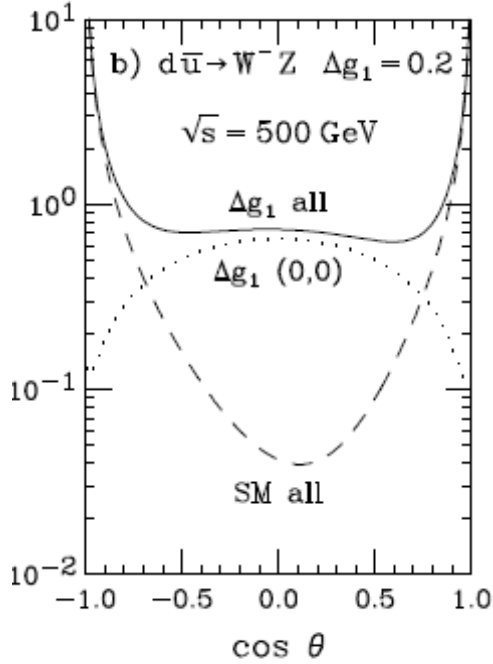

Figure 8: Angular distribution of SM (long dashed) and $Z_L W_L$ (dotted) events at 500 GeV. The full curve shows the resulting distribution with an anomalous coupling.

## Anomalous couplings

Mixing between Z, Zkk and Z' can generate anomalies. As previously one takes into account the weakness of the mixing angles to simplify the results.

1/ WW mode

For the SM Z exchange term, one has Z-Zkk and Z-Z' mixing at both vertices. At the Z->WW vertex Z-Zkk mixing generates a coupling anomaly which goes like:

$$\Delta \kappa_{ZWW} = c_{01} - 2 s_{01} s_{0L} - 1$$

Similarly Z-Z' mixing gives:

$$\Delta \kappa_{ZWW} = c_{01X} c_{0R}^2 + s_{0L}^2 - 2 s_{01X} s_{0L} - 1$$

With our parameters the total effect is $\Delta \kappa_{Zww}$=−0.00089+0.00011=-0.00078

At the Zqq vertex one has:

$$\Delta \kappa_{qqZ} = (c_1 s_{01} + s_{X01} s_1) Q(cqL) + \frac{g_{Z'}}{g_Z} \frac{\frac{I_{3R}^q}{2} - Y_q s'^2}{\frac{I_{3L}^q}{2} - Q_q s_W^2} (-s_{01} s_1 + c_1 s_{01X}) Q'(cqL) + c_1 - 1$$

where I3R and I3L were defined in the Appendix section on $W_L W_L$. Note that the predominant contribution comes from the term c1-1, hence is negative.

At first order: $\Delta\kappa_Z = \Delta\kappa_{Zww} + \Delta\kappa_{Zqq}$

This expression is derived using table 9 from ref8 noting that our isospin assignment is different. This gives $\Delta\kappa_Z$ **=-0.041** for u quarks and $\Delta\kappa_Z$ **=-0.049** for d quarks. These values therefore are predominant and, as shown in figure 1, induce a large effect.

2/ ZW mode

At the W->ZW vertex one has for the mixing terms:
$$\Delta g_{1WZW} = c_{01} c_{0L}^2 - 1 + c_{01X} c_{0R}^2 - 1 \text{ giving } \Delta g_{1Z} = -0.0025.$$

At the Wqq vertex one has:
$$\Delta g_{1qqZ} = c_{0R} - 1 - Q(c)\,(s_1 s_{0R} + c_1 s_{0L}) + Q'(c)\,(c_1 s_{0R} - s_1 s_{0L})\frac{g_R}{g_L} \quad \text{giving } \Delta g_{1Z} = -0.067 \text{ due,}$$

essentially, to the last term. The sum of these two contributions gives **$\Delta g_{1Z}$=-0.069**.

From this, one concludes that anomalous couplings can provide very good discrimination between the various scenarios allowed by RS models. If they turn out to be large, say at the % level, this would be an indication of the presence of substantial couplings of the light quarks to the KK resonances. As in figure 1, one could observe a precautious rise in the rate of $W_L W_L$.

# Breit Wigner curves

Near the resonances, the KK propagators need to be replaced by Breit Wigner curves of the type:

$$\frac{1}{s - m_{KK}^2} \rightarrow \frac{1}{s - m_{KK}^2 + iks} \quad \text{where } k = \frac{\Gamma_{KK}}{m_{KK}} \text{ is a constant defined at the resonance.}$$

## Akk/Zkk/Z'

Assuming, as in ref2, that tR has the highest overlap function Q(ctR)~6, the top contribution to Akk and Zkk are given by:

$$k_Z = \frac{Nc\left[Q(ctR)(-2s_W^2/3)g\right]^2}{48\pi} \qquad k_A = \frac{Nc\left[Q(tR)(-2/3)e\right]^2}{48\pi}$$

For Z', one has:

$$k_{Z'} = \frac{Nc[(I_{3R} - 2s'^2/3)Q'(tR)g_{Z'}]^2}{48\pi}$$

assuming that tR belongs to an isodoublet with I3R=1/2 (ref1). If tR, as in ref8, behaves as an isosinglet then tL mainly contributes with Q(tL)~1.9 from ref1. In the next table one assumes that tR is not a singlet and take Q(tR)~Q'(tR)=6 .
Generally speaking, one is dealing with large widths, in particular for what concerns the Z' resonance.

Note that when one increases Mkk, mixing angles will decrease and one needs to adjust above ratios for what concerns WW and ZW.

| Res | Akk | Zkk | Z' | Wkk | W' |
|---|---|---|---|---|---|
| Γ/M from Zh/Wh | 0 | 0.01 | 0.35 | ~0 | 0.35 |
| Γ/M from $W_L W_L$/ $Z_L W_L$ | 0.04 | 0.115 | 0.44 | 0.36 | 0.44 |
| Γ/M from ttbar/tbbar | 0.018 | 0.01 | 0.58 | 0.02 | ~0 |
| Γ/M tot | 0.06 | 0.134 | 1.4 | 0.36 | 0.79 |

## Wkk/W'

1/ Wh

As shown in ref 9, Wkk and W' couplings are very similar to Zkk and Z' implying similar conclusions. Zkk couples weakly to Wh while W' will have a large width.

2/ ZW

One finds that Qwkk=−0.025 and Qw'==−0.045 to be compared to Qzkk==− 0.01 and Qz'==−0.05

3/ tbbar

In ref8, in the minimal scenario where tR is an isosinglet, Wkk and W' only interacts with tLbL. Since in ref1 one assumes that tL and bL have overlap integral Q(tL)=1.9, it follows that both the Wkk and W' will receive a low contribution to their widths due to fermions.

$$k = \frac{Nc \left[ c_1 Q(tL) g_L \right]^2}{96\pi}$$

For W', one has:

$$k' = \frac{Nc [s_1 Q'(tL) g_L]^2}{96\pi}$$

This picture could of course be radically changed if tR belongs to a multiplet with new heavy quarks as explained in ref8. Note also that a quark with mass ~ 1 TeV, hardly detectable at LHC, could still affect the width at the Wkk resonance. Then Wkk could have a larger width, similarly to Zkk. In this paper one uses the values from the table assuming only standard decays.